\documentclass[a4paper,11pt]{article}
\pdfoutput=1
\usepackage{jcappub}
\usepackage[T1]{fontenc}
\usepackage{lmodern}
\usepackage{amsthm}
\usepackage{booktabs,array}
\usepackage{microtype}
\theoremstyle{plain}
\newtheorem{lemma}{Lemma}
\newtheorem{proposition}{Proposition}
\newtheorem{corollary}{Corollary}
\theoremstyle{remark}
\newtheorem{remark}{Remark}
\newcommand{\dd}{\mathrm d}
\newcommand{\cs}{c_{\mathrm s}^{2}}
\newcommand{\jump}[1]{\left[#1\right]}
\newcommand{\Hphase}{\mathrm H}
\newcommand{\Qphase}{\mathrm Q}
\newcommand{\rhoref}{\rho_{\mathrm{ref}}}
\title{One gravitational function across matter phases: compatibility and
interface constraints in static \texorpdfstring{\(f(R,T)\)}{f(R,T)} gravity}

\author{Samuel Isidoro dos Santos J\'unior}
\author{and Rafael C.\ R.\ de Lima}

\affiliation{Departamento de F\'isica, Universidade do Estado de Santa
Catarina (UDESC),\\ Joinville, SC, 89219-710, Brazil}

\emailAdd{samuel.isidoro.santos@gmail.com}
\emailAdd{rafael.lima@udesc.br}

\abstract{Conservative \(f(R,T)=R+h(T)\) gravity fixes \(h\) by demanding that the
matter stress tensor be conserved, and the reconstruction is normally
carried out for one equation of state at a time. We ask whether one
function \(h(T)\) can serve two matter phases separated by a sharp
transition. The analysis is static and local: spherically symmetric
backgrounds, the \(\mathcal L_m=\rho\) closure, cold barotropic perfect
fluids, and a Maxwell interface without a surface layer. A common \(h\)
must satisfy two conditions that single-phase reconstructions do not
impose. It must solve both reconstruction equations at every trace value
occupied by both phases, and it must make the effective normal pressure
continuous across the interface; equality of material pressure and
chemical potential does not secure the latter. For a hadronic polytrope
(\(K>0\), \(\Gamma>1\)) and an affine quark equation of state, an open
overlap of the occupied trace ranges forces \(h\) to be constant on both.
When the ranges are separated, both conditions can be met by a
nonconstant \(C^2\) function, which we construct explicitly at a single
interface. Two further results bound this local construction. If the
same polytrope extends to zero density, regularity of \(h\) at \(T=0\)
removes its nonconstant part there; and wherever \(h'(T)\neq0\), separate
conservation forbids any change in the density of a fluid element.}

\keywords{modified gravity, neutron stars}

\begin{document}
\maketitle
\flushbottom

\section{Introduction}
\label{sec:intro}
In \(f(R,T)\) gravity the gravitational action depends on the trace
\(T\) of the matter stress tensor \cite{Harko2011}. This coupling
generally spoils separate conservation of \(T_{\mu\nu}\). Conservative
variants restore it by choosing the trace-dependent part of the action
accordingly, first for homogeneous cosmologies
\cite{Alvarenga2013,Chakraborty2013}. For a static star, conservation
becomes a differential equation that reconstructs \(h(T)\) in
\(f(R,T)=R+h(T)\) from the equation of state \cite{Santos2019}. The
construction has been applied to neutron stars \cite{Santos2019,Pretel2021}
and to strange quark stars \cite{Carvalho2020}, and the reconstructed
function depends on the equation of state \cite{Santos2019,Lobato2026}.
A hybrid star, with a hadronic envelope and a quark core separated by a
first-order transition, places both kinds of matter in one spacetime.
Because \(h\) belongs to the gravitational action, it must then be a
single function of \(T\), evaluated in both phases. Reference
\cite{Pretel2021} identified this case, with its interface
discontinuity, as requiring junction conditions that were left for
future work.

Solving the reconstruction equation phase by phase does not provide such
a function. Each solution is defined only on the trace values its phase
occupies, and only up to integration constants. Two obstacles appear when
the pieces are combined. First, distinct matter states can share a trace
value, and one function must then satisfy both reconstruction equations
there. Second, at a sharp transition the coexistence conditions concern
the matter: equal pressure and equal chemical potential. The metric,
however, responds to an effective stress that also contains \(h\) and
\(h'(T)\). Material coexistence does not by itself make that effective
stress continuous.

We therefore ask the following question. Given two equations of state,
does a single \(C^2\) function \(h(T)\) exist that conserves the stress
tensor of each phase in bulk and admits a shell-free interface at the
Maxwell point? We answer it for static, spherically symmetric backgrounds
under the \(\mathcal L_m=\rho\) closure of Ref.~\cite{Santos2019}.
Section~\ref{sec:model} states the hypotheses precisely.

The answer consists of two necessary conditions and their consequences
for a hadronic polytrope paired with an affine quark phase. Function
compatibility on shared traces (Lemma~\ref{lem:shared}) forces \(h\) to
be constant whenever the two occupied trace ranges overlap on an open
interval (Proposition~\ref{prop:obstruction}). Geometric compatibility at
the interface adds one scalar condition, continuity of \(q(\rho+p)-h/2\)
with \(q=h'(T)\) (Proposition~\ref{prop:interface}); material equilibrium
leaves this quantity undetermined. When the trace ranges are separated,
both conditions can be met by a nonconstant function
(Proposition~\ref{prop:construction}), and we construct one explicitly.
Propositions~\ref{prop:surface} and~\ref{prop:dynamics} then establish
what this local construction cannot provide: a nonconstant \(h\) along a
polytropic envelope that reaches zero density, and compressible time
dependence.

Junction conditions in \(f(R,T)\) gravity generally require a continuous
trace, but not when \(f_R\) is constant
\cite{Rosa2021,RosaRubiera2022,Rosa2024}. We derive the interface
condition directly in that subclass and for the closure used here. The
meaning of a perfect-fluid matter Lagrangian and of its metric variation
is itself under discussion
\cite{Bertini2023,Haghani2024,Ky2026,CarvalhoLobato2026}; every statement
below is conditional on the closure fixed in Section~\ref{sec:model}.
Section~\ref{sec:previous} states what is specific to this work.

Section~\ref{sec:model} fixes the model, hypotheses, and terminology.
Section~\ref{sec:shared} treats shared trace values and proves the
obstruction. Section~\ref{sec:interface} derives the interface condition
and shows why Maxwell equilibrium does not imply it.
Section~\ref{sec:construction} constructs a common function for
separated ranges, and Section~\ref{sec:scope} delimits its physical
scope. Section~\ref{sec:conclusions} answers the opening question.

\section{Model, assumptions, and terminology}
\label{sec:model}
This section fixes once the equations and hypotheses under which the
question is posed; later statements cite the hypotheses they use.

\subsection{Field equations and closure}
We use signature \((+---)\), \(G=c=1\), and
\begin{equation}
 T_{\mu\nu}=(\rho+p)u_\mu u_\nu-pg_{\mu\nu},\qquad
 u^\mu u_\mu=1,\qquad T=\rho-3p,
\end{equation}
where \(\rho\) is the total energy density. We adopt the
\(\mathcal L_m=\rho\) closure of the original static construction
\cite{Santos2019}, \(\Theta_{\mu\nu}=-2T_{\mu\nu}+\rho g_{\mu\nu}\). This
is a stipulated metric-variation prescription, not a result derived from
a fluid action: derivations from the full fluid actions of Brown and
Schutz \cite{Brown1993,Schutz1970} give different trace-coupling terms
\cite{Boehmer2025}, and Section~\ref{sec:previous} discusses what this
implies for the present results. Writing \(q(T)=h_T(T)\), the
field equations of \(f(R,T)=R+h(T)\) become
\begin{equation}
 G_{\mu\nu}=8\pi T_{\mu\nu}+\frac{h}{2}g_{\mu\nu}
                  +q(T)(T_{\mu\nu}-\rho g_{\mu\nu}).
 \label{eq:field}
\end{equation}
All component expressions and junction results below follow from
Eq.~\eqref{eq:field}; Appendix~\ref{app:sign} gives a component-level sign
audit.

\subsection{Assumptions}
Each result below lists which of the following hypotheses it uses.
\begin{itemize}
 \item[(A1)] \emph{One fixed function.} Equation~\eqref{eq:field} holds
 with \(h\in C^2\) on a real interval containing every occupied trace
 range and the intervals between them. Once chosen, the same \(h\) is
 evaluated in every phase and every configuration.
 \item[(A2)] \emph{Separate conservation.} The material stress tensor
 obeys \(\nabla^\mu T_{\mu\nu}=0\) in the bulk of each phase and, in the
 weak sense, across interfaces. This is an additional restriction:
 diffeomorphism invariance guarantees conservation of the total source on
 the right-hand side of Eq.~\eqref{eq:field}, not of \(T_{\mu\nu}\) by
 itself \cite{BarrosSa2026}.
 \item[(A3)] \emph{Regular barotropic branches.} Each phase is a cold
 perfect fluid with a \(C^1\) barotrope \(p(\rho)\),
 \(w\equiv\rho+p>0\), and \(0\le\cs=\dd p/\dd\rho\le1\). It occupies a
 connected density interval of positive length, called its branch; the
 corresponding set of trace values is its occupied trace range.
 \item[(A4)] \emph{Static background.} The spacetime is static and
 spherically symmetric, and the fluid is at rest.
 \item[(A5)] \emph{Sharp Maxwell interface.} The phases meet at a static
 comoving hypersurface with no flux across it, no surface layer of
 matter, and no surface tension. The transition is a cold, locally
 neutral Maxwell transition with one conserved baryon number. Globally
 neutral mixed phases, which carry an additional independent chemical
 potential, lie outside this construction \cite{Glendenning1992}.
\end{itemize}
Constant-trace phases and branches through a trace turning point violate
the invertibility used in Section~\ref{sec:shared}; we treat them
separately there. Proposition~\ref{prop:dynamics} drops (A4).

We distinguish three notions throughout. \emph{Material equilibrium}
refers to the matter alone: bulk conservation (A2) and, at the interface,
equal pressure and chemical potential. \emph{Function compatibility}
requires one \(h(T)\) whose restriction to each occupied trace range
solves that phase's reconstruction equation; where traces coincide, the
two phases must see the same \(h\), \(q\), and \(q_T\). \emph{Geometric
compatibility} requires the metric to join across the interface without
a surface layer.

For numerical examples we introduce an arbitrary reference density
\(\rhoref>0\), define \(\bar K=K\rhoref^{\Gamma-1}\), and write \(\bar\rho=\rho/\rhoref\), \(\bar p=p/\rhoref\),
\(\bar T=T/\rhoref\), and \(\bar h=h/\rhoref\); then
\(q=\dd\bar h/\dd\bar T\) is dimensionless. Dimensions in geometrized
units are listed in Appendix~\ref{app:formulas}.

\subsection{Conservation constraint and single-phase reconstruction}
Imposing (A2) in a smooth region and taking the divergence of
Eq.~\eqref{eq:field} gives
\begin{equation}
 (T_{\mu\nu}-\rho g_{\mu\nu})\nabla^\mu q
        -\frac{q}{2}\nabla_\nu(\rho+3p)=0.
 \label{eq:covconstraint}
\end{equation}
No division by \(q\) or by \(8\pi+q\) is required. Under (A4) the only
nontrivial component is radial,
\begin{equation}
 w\frac{\dd q}{\dd r}+\frac{q}{2}\frac{\dd(\rho+3p)}{\dd r}=0,
 \label{eq:radial}
\end{equation}
which is the static conservation condition of Ref.~\cite{Santos2019}. On a
branch where the density varies, Eq.~\eqref{eq:radial} becomes an
equation in \(\rho\), with solution
\begin{align}
 \frac{\dd q}{\dd\rho}&=-\frac{1+3\cs}{2w}\,q,
 \label{eq:q-rho}\\
 q(\rho)&=C\exp\left[-\int_{\rho_a}^{\rho}
           \frac{1+3\cs(z)}{2[z+p(z)]}\,\dd z\right],
 \label{eq:qsol}\\
 h(\rho)&=B+\int_{\rho_a}^{\rho}q(z)[1-3\cs(z)]\,\dd z.
 \label{eq:hsol}
\end{align}
The normalization \(C\) and the additive constant \(B\) are independent
integration data. The branch \(C=0\), with \(q=0\) and constant \(h\),
must not be lost by writing \(\ln q\). For \(C\neq0\) the exponential is
positive, so \(q\) cannot cross zero on a connected branch; if \(q\)
vanishes at one point of a branch, it vanishes on the whole branch. Where
\(T_\rho=1-3\cs\neq0\), the same equation reads, as a function of the
trace,
\begin{equation}
 \frac{\dd q}{\dd T}=-F(T)\,q,\qquad
 F(T)=\frac{1+3\cs}{2w\,(1-3\cs)},
 \label{eq:F}
\end{equation}
with matter quantities evaluated on the inverse branch \(\rho(T)\).

A single-phase reconstruction therefore determines \(h\) only on the
trace values its phase occupies, and only up to the pair \((C,B)\).
Whether two such pieces belong to one function is decided by the
conditions of the next two sections.

\section{Function compatibility on shared trace values}
\label{sec:shared}
When two phases occupy the same trace value, one function \(h\) must
serve both. This section gives the criterion for that and applies it to
the polytrope--affine pair.

\subsection{General criterion}
\begin{lemma}[Shared-trace criterion]
\label{lem:shared}
Assume (A1)--(A4). Let two branches have a connected open trace overlap
\(J\), on which \(1-3{\cs}_i\neq0\). A common nonzero \(q\) solves both
reconstruction equations on \(J\) if and only if \(F_1=F_2\) on \(J\) and
the two normalizations agree at one point; a common \(h\) further
requires one common additive constant. If \(F_1\neq F_2\) at a single
point of \(J\), then \(q=0\) on both branches, and \(h\) takes one and the
same constant value on both.
\end{lemma}
\begin{proof}
Subtracting the two equations \(q_T=-F_iq\) gives \((F_1-F_2)q=0\) on
\(J\). If the coefficients differ at one point, \(q\) vanishes there.
Uniqueness for the linear equation~\eqref{eq:q-rho} propagates the zero
along each connected branch, so \(h\) is constant on each; the two
constants coincide because the branches share a trace value. If the
coefficients agree, uniqueness gives the same \(q\) after one common
normalization, and integrating \(h_T=q\) gives the same \(h\) after one
common additive constant.
\end{proof}

Equality at a single trace value is weaker than an open overlap. At an
isolated contact, a \(C^2\) function must have matching \(h\), \(q\), and
\(q_T\). Unless \(q=0\) there, this requires equal \(F_i\) at the contact,
but not on a nonexistent open interval.

A branch may also fold in trace. Every repeated trace value must then
yield the same \(h\), \(q\), and \(q_T\); one may not pick a different
inverse branch at each radius without testing these equalities.
\begin{lemma}[Trace turning point]
\label{lem:turning}
Assume (A1)--(A4). If a branch contains a density \(\rho_*\) with
\(\cs(\rho_*)=1/3\), then \(q=0\) on that branch.
\end{lemma}
\begin{proof}
By the chain rule, \(\dd q/\dd\rho=h_{TT}(1-3\cs)=0\) at \(\rho_*\), since
\(h_{TT}\) is finite by (A1). Equation~\eqref{eq:q-rho} gives
\(\dd q/\dd\rho=-q/w\) there. Hence \(q(\rho_*)=0\), and uniqueness propagates the zero along the
branch.
\end{proof}

Lemma~\ref{lem:turning} applies to any occupied branch that reaches
\(\cs=1/3\). Realistic equations of state often do: supporting
two-solar-mass stars with the hadronic equation of state known at low
density is in strong tension with \(\cs\le1/3\) \cite{Bedaque2015}. In a
large ensemble of equations of state constrained by nuclear theory,
perturbative QCD, and astrophysical data, about half produce
general-relativistic stars in which the trace is negative somewhere
\cite{Garibay2026}. If the same continuous branch extends from
\(\rho=0\), with \(p/\rho\to0\) as \(\rho\to0^+\), then \(T(0)=0\)
and \(T>0\) at sufficiently low density. A negative trace at a higher
density requires an interior maximum of \(T\), where \(\cs=1/3\) and
Lemma~\ref{lem:turning} applies. A sign change on an arbitrary branch
starting at finite density does not suffice: an affine phase can cross
\(T=0\) with constant \(\cs\neq1/3\). Applications to realistic equations
of state must therefore test trace invertibility on each occupied phase
branch explicitly.

An affine phase with \(\cs=1/3\) has constant trace \(T_0\). Along any
portion where its density varies, \(\dd q/\dd r=h_{TT}\,\dd T/\dd r=0\)
while \(\dd(\rho+3p)/\dd r=2\,\dd\rho/\dd r\), so Eq.~\eqref{eq:radial}
gives
\begin{equation}
 q(T_0)=0.\label{eq:constanttrace}
\end{equation}
Such a phase samples only the value \(h(T_0)\), which acts as a constant
geometric contribution; nothing follows for \(q\) at other trace values.

\subsection{Polytropic and affine branches}
For the hadronic polytrope \(p_\Hphase=K\rho^\Gamma\) with \(K>0\) and
\(\Gamma>1\), set \(x=K\rho^{\Gamma-1}\), so that \(\cs=\Gamma x\),
\(w=\rho(1+x)\), and \(T_\Hphase=\rho(1-3x)\). Equation~\eqref{eq:qsol}
gives
\begin{equation}
 q_\Hphase=C_\Hphase\,\rho^{-1/2}
       \left(1+K\rho^{\Gamma-1}\right)^{-\frac{3\Gamma-1}{2(\Gamma-1)}},
 \label{eq:qH}
\end{equation}
with fractional powers taken on positive \(\rho\). The trace turns at
\begin{equation}
 \rho_*=(3\Gamma K)^{-1/(\Gamma-1)},
 \label{eq:rhostar}
\end{equation}
and the reconstruction coefficient on a branch that avoids \(\rho_*\) is
\begin{equation}
 F_\Hphase=\frac{1+3\Gamma x}{2\rho\,(1+x)(1-3\Gamma x)}.
 \label{eq:FH}
\end{equation}
Although \(\cs=1/3\) is causal, Lemma~\ref{lem:turning} prevents a
nonzero reconstruction from crossing \(\rho_*\).

For the affine quark phase \(p_\Qphase=s(\rho-\rho_0)\) with
\(0<s\leq1\) and \(s\neq1/3\), let \(w_\Qphase=(1+s)\rho-s\rho_0>0\) and
\(T_\Qphase=(1-3s)\rho+3s\rho_0\). Then
\begin{equation}
 q_\Qphase=C_\Qphase\,w_\Qphase^{-(1+3s)/(2+2s)},
 \label{eq:qQ}
\end{equation}
and, expressing \(w_\Qphase\) through \(T\),
\begin{equation}
 F_\Qphase=\frac{1+3s}{2\left[(1+s)T-4s\rho_0\right]}.
 \label{eq:FQ}
\end{equation}
The denominator equals \(2(1-3s)w_\Qphase\neq0\). The primitives \(h_i\),
including the logarithmic case \(s=1\), are collected in
Appendix~\ref{app:formulas}.

Dense-matter studies follow the trace across such transitions through
the normalized trace anomaly \(\Delta=1/3-p/\rho=T/(3\rho)\), which
changes sharply near the Maxwell point of twin-star models and can become
negative \cite{Jimenez2024}. The argument of \(h\), however, is the
unnormalized trace. At the Maxwell point itself \(\jump{T}=\jump{\rho}>0\)
at equal pressure, and features of \(\Delta\) do not translate directly
into features of \(T\).

\subsection{Obstruction for overlapping trace ranges}
\begin{proposition}[Polytrope--affine obstruction]
\label{prop:obstruction}
Assume (A1)--(A4) for a polytropic branch with \(K>0\), \(\Gamma>1\), and
an affine branch with \(0<s\leq1\). If their occupied trace ranges overlap
on a nonempty open interval, every common \(h\) is constant on both
occupied ranges.
\end{proposition}
\begin{proof}
If the polytropic branch contains \(\rho_*\), Lemma~\ref{lem:turning}
gives \(q=0\) on it. Then \(q\) vanishes at a trace value occupied by the
affine branch, hence on that whole branch by Eq.~\eqref{eq:qsol}, and the
two constant values of \(h\) coincide. If \(s=1/3\), the affine
trace range is a single point and no open overlap exists. Otherwise both
branches are invertible on the overlap. Equality \(F_\Hphase=F_\Qphase\)
at the hadronic trace \(T=\rho(1-3x)\) would require
\begin{equation}
 0=P(\rho)=(1+3\Gamma x)\left[(1+s)\rho\,(1-3x)-4s\rho_0\right]
             -(1+3s)\rho\,(1+x)(1-3\Gamma x).
 \label{eq:polyobstruction}
\end{equation}
With \(x=K\rho^{\Gamma-1}\), \(P\) is a finite sum of real powers of
\(\rho\). The largest exponent is \(2\Gamma-1\), with coefficient
\(-6\Gamma K^2\neq0\); the remaining exponents belong to
\(\{\Gamma,1,\Gamma-1,0\}\) and are strictly smaller for \(\Gamma>1\).
Powers with distinct real exponents are linearly independent on any
open interval of \(\rho>0\), so \(P\) cannot vanish on an open density
interval, and \(F_\Hphase\neq F_\Qphase\) somewhere on the overlap.
Lemma~\ref{lem:shared} then gives \(q=0\) on both branches and a common
constant \(h\).
\end{proof}

The comparison is made at equal trace, which generally corresponds to
different densities; the two equations of state need not agree at equal
\(\rho\). A constant \(h\) is general relativity with a geometric
constant, and pure general relativity requires \(h=0\). Assigning
\(h(0)=0\) does not by itself fix that constant on an occupied range
separated from \(T=0\).

\subsection{Illustration}
\label{sec:illustration}
We use synthetic data in units of \(\rhoref\), not an astrophysical fit.
The hadronic branch is \(\bar p_\Hphase=\bar\rho^{\,2}/10\) on
\(\bar\rho_\Hphase\in[1/2,1]\), with trace range \([0.425,0.7]\) and
turning density \(\bar\rho_*=5/3\) outside the branch. Two affine
candidates coexist with it at the same material state,
\begin{equation}
 (\bar\rho_\Hphase,\bar\rho_\Qphase,\bar p_t)=(1,2,1/10),
 \label{eq:coexistence}
\end{equation}
namely a stiff phase with \(s=3/5\), \(\bar\rho_0=11/6\), and a soft phase with
\(s=1/5\), \(\bar\rho_0=3/2\). Their Maxwell data also coincide
(Section~\ref{sec:maxwell}).

\begin{figure}[tbp]
\centering
\includegraphics[width=\textwidth]{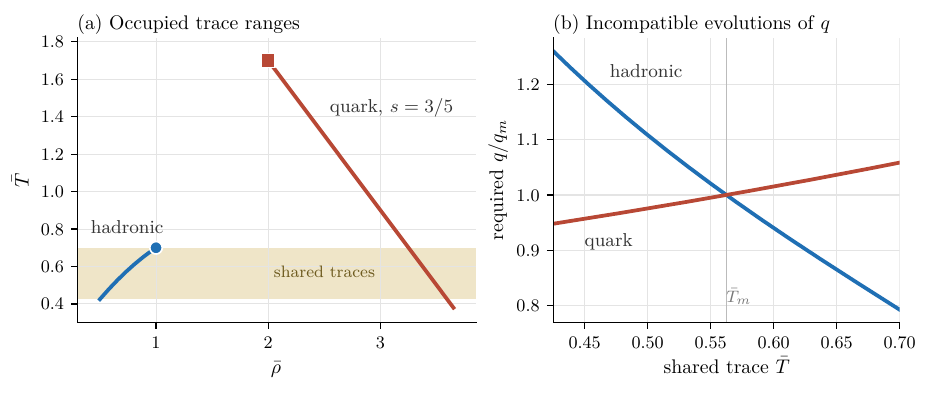}
\caption{Function compatibility fails on shared trace values. (a)~Trace
along the hadronic branch and the stiff affine branch (\(s=3/5\),
\(\bar\rho_0=11/6\)); the shaded band marks the trace values
occupied by both. The circle and the square mark the
coexistence states of Eq.~\eqref{eq:coexistence}. (b)~The function
\(q(\bar T)\) that each phase requires on the overlap, both normalized to
\(q_m\) at the midpoint \(\bar T_m\). The hadronic solution decreases
and the quark solution increases; normalizing at any other point gives
the same opposite trends, so only \(q=0\) satisfies both. These are
constitutive curves, not radial profiles.}
\label{fig:shared}
\end{figure}

The stiff phase, occupying \(\bar\rho_\Qphase\in[2,3.65]\), has a trace
that decreases from \(1.7\) to \(0.38\) and therefore covers the whole
hadronic trace range. Figure~\ref{fig:shared} shows the result. On the
shared interval \(F_\Hphase\geq1.63/\rhoref>0\) while
\(F_\Qphase\leq-0.376/\rhoref<0\): a nonzero \(q\) would have to decrease
with \(T\) according to the hadronic equation and increase according to
the quark equation. By Proposition~\ref{prop:obstruction}, only a
constant \(h\) serves both. The soft phase, on
\(\bar\rho_\Qphase\in[2,2.5]\), occupies the trace range \([1.7,1.9]\),
disjoint from the hadronic one; Sections~\ref{sec:interface}
and~\ref{sec:construction} use it.

For this pair, a nonconstant common function therefore requires disjoint
occupied trace ranges. Disjointness removes the shared-trace condition
altogether, so any further restriction on \(h\) must come from the
interface.

\section{Interface: material equilibrium does not fix the traction}
\label{sec:interface}
If the occupied trace ranges are disjoint, Section~\ref{sec:shared}
imposes nothing, and the matching might seem complete once material
equilibrium holds at the transition. This section shows that it is not.

\subsection{Material equilibrium}
\label{sec:maxwell}
Under (A5) each phase obeys the cold first law
\begin{equation}
 \dd\rho=\frac{\rho+p}{n}\dd n,\qquad
 \mu=\frac{\rho+p}{n},\qquad \dd p=n\,\dd\mu,
 \label{eq:firstlaw}
\end{equation}
which determines the baryon densities \(n_i\) up to normalizations
\(N_i>0\) (Appendix~\ref{app:formulas}). The interface conditions are
\begin{equation}
 p_\Hphase=p_\Qphase=p_t,\qquad
 \mu_\Hphase=\mu_\Qphase=\mu_t,\qquad
 \rho_\Qphase=\rho_0+\frac{p_t}{s}>
 \rho_\Hphase=(p_t/K)^{1/\Gamma}.
 \label{eq:Maxwell}
\end{equation}
For a microphysical equation of state the \(N_i\) are fixed before
coexistence is located. In our synthetic pairs we instead choose them so
that \(\bar n_\Hphase=1\) and \(\bar n_\Qphase=21/11\) at the state
\eqref{eq:coexistence}, with a separately fixed baryon-density scale; then
\(\bar\mu_t=11/10\) for both affine candidates. Since \(\dd\mu/\dd p=1/n\)
and \(n_\Qphase>n_\Hphase\) at coexistence, the quark phase has the lower
chemical potential just above \(p_t\) and the hadronic phase just below
(Fig.~\ref{fig:interface}a). The affine parameterization is related to
constant-sound-speed hybrid matter \cite{Alford2013}; we address
conservation and junction consistency, not the stability phase diagram
studied there.

\subsection{Effective stress}
Equation~\eqref{eq:field} can be written as
\(G_{\mu\nu}=8\pi T^{\rm eff}_{\mu\nu}\), with
\begin{equation}
 \rho_{\rm eff}=\rho+\frac{h}{16\pi},\qquad
 p_{\rm eff}=p+\frac{q\,(\rho+p)-h/2}{8\pi}.
 \label{eq:effective}
\end{equation}
We call the numerator of the pressure correction
\begin{equation}
 A=q\,(\rho+p)-\frac{h}{2}.
 \label{eq:A}
\end{equation}
Substituting Eqs.~\eqref{eq:q-rho} and~\eqref{eq:hsol} gives, on a
conservative branch,
\begin{equation}
 \frac{\dd A}{\dd\rho}=q\,\cs,
 \qquad \frac{\dd A}{\dd p}=q\quad(\cs\neq0).
 \label{eq:Aidentity}
\end{equation}
These identities hold within a branch. Integrating \(\dd A/\dd p\) across
a density jump is not legitimate, because no occupied barotropic path
spans the jump.

\subsection{Geometric compatibility}
Let \(\Sigma\) be the interface, with unit spacelike normal \(\ell^\mu\),
\(\ell^\mu\ell_\mu=-1\), and \(u^\mu\ell_\mu=0\). We write jumps as
\(\jump{X}=X_\Qphase-X_\Hphase\); their vanishing does not depend on the
orientation of \(\ell^\mu\).
\begin{proposition}[Interface traction condition]
\label{prop:interface}
Assume (A1), (A2), and (A5). A weak solution of Eq.~\eqref{eq:field}
across \(\Sigma\) requires the Darmois conditions
\cite{Darmois1927,Israel1966}
\(\jump{\gamma_{ab}}=\jump{\mathcal K_{ab}}=0\) for the induced metric
and extrinsic curvature, together with \(\jump{p}=0\) and
\begin{equation}
 \boxed{\jump{A}=0,\qquad
 h(T_\Qphase)-h(T_\Hphase)
 =2\left\{q(T_\Qphase)(\rho_\Qphase+p_t)
       -q(T_\Hphase)(\rho_\Hphase+p_t)\right\},}
 \label{eq:interface}
\end{equation}
where all quantities are evaluated at the coexistence states.
\end{proposition}
\begin{proof}
By (A5), \(T_{\mu\nu}\) has no singular part on \(\Sigma\); by (A1),
\(h(T)\) and \(q(T)\) are bounded. Hence \(T^{\rm eff}_{\mu\nu}\) is
bounded, and the singular part of \(G_{\mu\nu}\), which is built from
\(\jump{\mathcal K_{ab}}\), must vanish. Since \(f_R=1\), the field
equations contain no derivatives of \(h\) and no \(\nabla\nabla f_R\)
terms that could contribute further singular parts. The Hamiltonian and
momentum constraints on \(\Sigma\) express
\(G_{\mu\nu}\ell^\mu\ell^\nu\) through \(\gamma_{ab}\) and
\(\mathcal K_{ab}\), so it is continuous; for a comoving interface
\(T^{\rm eff}_{\mu\nu}\ell^\mu\ell^\nu=p_{\rm eff}\), and therefore
\(\jump{p_{\rm eff}}=0\). Separately, by (A2) the singular part of
\(\nabla^\mu T_{\mu\nu}\), namely \(\ell^\mu\jump{T_{\mu\nu}}
=-\jump{p}\ell_\nu\), must vanish. Subtracting the two continuity
conditions with Eq.~\eqref{eq:effective} gives \(\jump{A}=0\).
\end{proof}

The trace generally jumps, \(\jump{T}=\jump{\rho}\) at equal pressure.
This is admissible here: \(h(T)\) and \(q(T)\) are bounded piecewise
functions, the field equations contain no derivatives of them, and no
product of a step with a delta function arises. This agrees with the
general junction analyses, which permit \(\jump{T}\neq0\) when \(f_R\) is
constant \cite{Rosa2021,RosaRubiera2022,Rosa2024}. The weak divergence
must be taken of the full effective tensor; the smooth chain rule should
not be applied at \(\Sigma\). In the limit \(h=q=0\), Eq.~\eqref{eq:interface}
holds automatically and the general-relativistic junction is recovered.

\subsection{What material equilibrium leaves open}
On each occupied range, bulk conservation fixes \(h\) up to the pair
\((C_i,B_i)\) of Eqs.~\eqref{eq:qsol}--\eqref{eq:hsol}. The Maxwell
conditions~\eqref{eq:Maxwell} involve only the matter and contain none of
these constants, whereas \(\jump{A}\) depends linearly on all four. Bulk
conservation together with material equilibrium therefore leaves
\(\jump{p_{\rm eff}}=\jump{A}/(8\pi)\) undetermined, and
Eq.~\eqref{eq:interface} is a genuinely additional condition.

\begin{figure}[tbp]
\centering
\includegraphics[width=\textwidth]{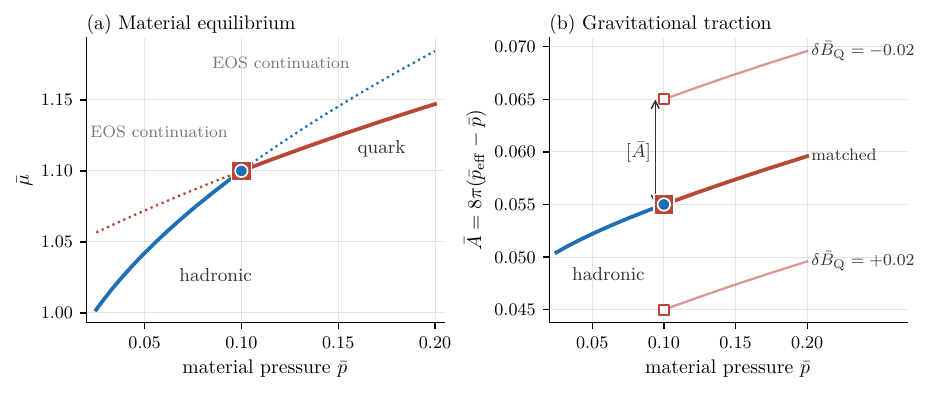}
\caption{Material equilibrium and gravitational traction at the same
interface (soft pair, \(s=1/5\), \(\bar\rho_0=3/2\)). (a)~Chemical
potential against material pressure. Solid curves are the occupied
branches; dotted curves continue each equation of state across the
transition. The curves cross once, at \(\bar p_t=1/10\), where
the circle and square mark the two coexistence states. (b)~Pressure
correction \(A=8\pi(p_{\rm eff}-p)\) on the occupied branches. With the
bulk data of Eq.~\eqref{eq:bulkdata} the two branches meet at
\(\bar p_t\), so \(\jump{p_{\rm eff}}=0\). Shifting only the additive
constant of the quark reconstruction by \(\delta\bar B_\Qphase=\pm0.02\)
(thin curves) leaves panel~(a) unchanged but opens a gap
\([\bar A]=-\delta\bar B_\Qphase/2\).}
\label{fig:interface}
\end{figure}

The soft pair of Section~\ref{sec:illustration} makes this explicit. Fix
the bulk data at coexistence as
\begin{equation}
 (q_\Hphase,q_\Qphase)=(1/20,1/20),\qquad
 (\bar h_\Hphase,\bar h_\Qphase)=(0,1/10),
 \label{eq:bulkdata}
\end{equation}
which determines both bulk functions (Appendix~\ref{app:formulas}). Then
\begin{equation}
 \bar A_\Hphase=\bar A_\Qphase=\frac{11}{200},\qquad
 \bar p_{\rm eff,t}=\frac{1}{10}+\frac{11}{1600\pi},
 \label{eq:Avalues}
\end{equation}
so the interface condition holds. Now shift the quark constant \(B_\Qphase\)
by \(\delta\bar B_\Qphase\), keeping \(q_\Qphase\) and the matter data
fixed. Bulk conservation and Maxwell equilibrium remain satisfied, but
\begin{equation}
 \jump{\bar p_{\rm eff}}=-\frac{\delta\bar B_\Qphase}{16\pi}.
 \label{eq:offsetresidual}
\end{equation}
The shifted quark function is an equally valid single-phase
reconstruction; it fails only the shell-free interface test.
Figure~\ref{fig:interface} displays both statements.

\subsection{Local geometric sufficiency}
Under (A4) the metric is
\begin{align}
 \dd s^2&=e^\phi\dd t^2-(1-2m/r)^{-1}\dd r^2-r^2\dd\Omega^2,\\
 m'&=4\pi r^2\rho_{\rm eff},\qquad
 \phi'=\frac{2(m+4\pi r^3p_{\rm eff})}{r(r-2m)},\qquad
 p'=-\frac{w\phi'}2.
 \label{eq:localTOV}
\end{align}
Given smooth bulk data with \(\cs>0\), finite coefficients, \(r_t>2m_t\),
a common \(m_t\), and Eq.~\eqref{eq:interface}, the two one-sided local
solutions share the induced metric and extrinsic curvature after a lapse
normalization; \(m'\) may jump. Together with \(\jump p=0\) this is a
local weak solution of Eq.~\eqref{eq:field}, and the angular equation
follows from the Bianchi identity and effective conservation.

Consequently, at each interface a common \(h\) must satisfy one scalar
relation between its values and slopes at the two coexistence traces,
Eq.~\eqref{eq:interface}, and material equilibrium supplies none of it.

\section{A common function for separated trace ranges}
\label{sec:construction}
The shared-trace and traction conditions together still leave room for
a nonconstant common function, as this section shows.

\begin{proposition}[Separated ranges]
\label{prop:construction}
Assume (A1)--(A5) for two phases whose occupied trace ranges are disjoint
compact intervals. On each corresponding compact density interval,
assume that \(p\) extends as a \(C^1\) function to a neighbourhood,
\(w>0\), and \(1-3\cs\neq0\), including at the endpoints. Let
\((C_\Hphase,B_\Hphase,C_\Qphase,B_\Qphase)\) be bulk data satisfying
Eq.~\eqref{eq:interface}. Then there is a function \(h\in C^2\) on the
smallest interval containing both ranges that coincides with the two
bulk reconstructions on their occupied ranges. Because
Eq.~\eqref{eq:interface} is one linear relation among four constants,
such bulk data form a three-parameter family admitting nonconstant
reconstructions.
\end{proposition}
\begin{proof}
The stated regularity and invertibility make each bulk reconstruction
\(C^2\) up to the endpoints of its occupied trace range. On each range it
fixes \(h\), \(q\), and
\(q_T=-Fq\) at the endpoint facing the gap. The gap has positive width,
and a quintic Hermite polynomial matches these six values, giving a
\(C^2\) extension. The shared-trace condition is void because the ranges
are disjoint, and Eq.~\eqref{eq:interface} holds by assumption. The
relation~\eqref{eq:interface} is linear in the four constants and fixes,
for instance, \(B_\Qphase\) once the others are chosen.
\end{proof}

The interior of the gap is not sampled by this configuration, so any
\(C^2\) extension with the same endpoint jets serves it equally. The
three parameters count bulk integration data, not this functional
freedom: adding any \(C^2\) function supported strictly inside the gap
preserves the matching, so the extensions form an infinite-dimensional
family. This freedom is part of designing one fixed function, not permission to retune
\(h\) while integrating a star; matter that occupies the gap in another
configuration would sample the chosen extension.

\begin{figure}[tbp]
\centering
\includegraphics[width=\textwidth]{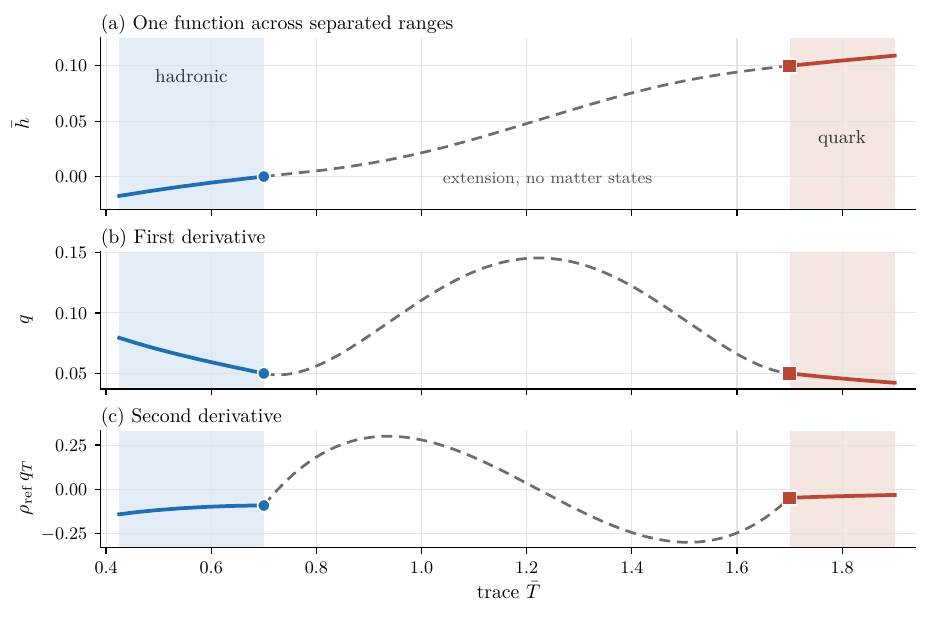}
\caption{A single \(C^2\) function serving both phases (soft pair, bulk
data of Eq.~\eqref{eq:bulkdata}). Panels show (a)~\(\bar h\),
(b)~\(q=\dd\bar h/\dd\bar T\), and (c)~\(\rhoref\,q_T\) against the trace.
Shaded bands mark the occupied trace ranges, where \(h\) is fixed by bulk
conservation; the dashed curve is the extension~\eqref{eq:bridge} across
the unoccupied gap, where no matter state of this configuration lies. The
circle and square mark the coexistence traces. Values, first, and second
derivatives match at both ends, and the traction condition holds there
with \(\bar A=11/200\).}
\label{fig:construction}
\end{figure}

For the soft pair with the bulk data~\eqref{eq:bulkdata}, the occupied
trace ranges are \([0.425,0.7]\) and \([1.7,1.9]\). With
\(x=\bar T-0.7\in[0,1]\) on the gap, an explicit extension is
\begin{equation}
 \bar h_{\rm gap}=\frac{x}{4620}\left(1486x^4-3875x^3+2830x^2-210x+231\right).
 \label{eq:bridge}
\end{equation}
It satisfies
\begin{equation}
 (\bar h,q,\rhoref q_T)\big|_{x=0}=(0,1/20,-1/11),\qquad
 (\bar h,q,\rhoref q_T)\big|_{x=1}=(1/10,1/20,-1/21),
 \label{eq:jets}
\end{equation}
where the second derivatives are those required by Eq.~\eqref{eq:F} at
the two coexistence states. Together with Eq.~\eqref{eq:Avalues}, this
gives one \(C^2\) function that satisfies both compatibility conditions.
Figure~\ref{fig:construction} shows it.

For the polytrope--affine pair the answer to the opening question thus
depends on the trace geometry: an open overlap admits only a constant
\(h\), whereas separated ranges admit a nonconstant common function that
satisfies both conditions at the interface. The construction is local,
and the next section establishes what that qualification excludes.

\section{Physical scope of the construction}
\label{sec:scope}
The construction of Section~\ref{sec:construction} is local. This
section identifies two parts of a hybrid star that lie outside it and
states each limitation as a result rather than a caveat.

\subsection{A polytropic envelope reaching zero density}
\label{sec:surface}
\begin{proposition}[Zero-density endpoint]
\label{prop:surface}
Assume (A1)--(A4) for a polytropic branch with \(K>0\), \(\Gamma>1\),
whose densities come arbitrarily close to zero, and suppose \(q\) is
bounded on a neighbourhood of \(T=0\) within the domain of \(h\); in
particular, \(h\in C^1\) on an interval containing \(T=0\) suffices.
Then \(C_\Hphase=0\): on the whole branch \(q=0\) and \(h=h(0)\), and
\(h=0\) there if the exterior is asymptotically flat.
\end{proposition}
\begin{proof}
If the branch contains \(\rho_*\), Lemma~\ref{lem:turning} already gives
\(q=0\). Otherwise, by Eq.~\eqref{eq:qH}, \(|q_\Hphase|\) is bounded
below by a positive multiple of \(|C_\Hphase|\rho^{-1/2}\) on any
low-density part of the branch, while \(T_\Hphase=\rho(1-3K\rho^{\Gamma-1})
\to0^+\) as \(\rho\to0^+\). Boundedness of \(q\) near \(T=0\) forces
\(C_\Hphase=0\), hence \(q=0\) on the branch. Then \(h\) is constant
there, and continuity gives the value \(h(0)\). A vacuum exterior has
\(G_{\mu\nu}=h(0)g_{\mu\nu}/2\), which is asymptotically flat only if
\(h(0)=0\).
\end{proof}

\begin{remark}
\label{rem:surface}
Continuity of \(h\) alone would not suffice. If \(C_\Hphase\neq0\), then
\(h-h(0)\simeq2C_\Hphase T^{1/2}\) remains continuous, and the effective
stress stays finite because \(qw\propto\rho^{1/2}\to0\). The obstruction
concerns the regularity of the function in the action. Without it, and
with \(h(0)=0\), the same \(h\) evaluated on dilute matter with
\(p/\rho\to0\), for which \(T\simeq\rho\to0^+\), would give
\(\rho_{\rm eff}-\rho\simeq C_\Hphase T^{1/2}/(8\pi)\), which dominates
\(\rho\). The proof uses only \(q\propto\rho^{-1/2}\). With \(y=p/\rho\),
Eq.~\eqref{eq:q-rho} gives
\begin{equation}
 \frac{\dd}{\dd\rho}\ln\left(|q|\rho^{1/2}\right)
 =\frac{y-3\cs}{2\rho\,(1+y)},
 \label{eq:surfacegeneral}
\end{equation}
whose right-hand side is integrable at \(\rho=0\) for any barotrope with
\(p(0)=0\), \(0\le\cs\le1\), and \(\int_0\cs\,\dd\rho/\rho<\infty\). For
such branches a nonzero \(q\) again behaves as \(\rho^{-1/2}\). The
condition \(p(0)=0\) is essential: if \(p\) tends to a positive constant,
\(w\) stays finite and \(q\) remains bounded.
\end{remark}

\begin{corollary}[Polytropic envelope]
\label{cor:envelope}
Assume (A1)--(A5) and the hypotheses of Proposition~\ref{prop:surface} for
a hadronic envelope that extends from the interface to a vacuum surface,
with an asymptotically flat exterior. Then \(h=0\) on the hadronic trace
range, and Eq.~\eqref{eq:interface} reduces to
\begin{equation}
 h(T_{\Qphase,t})=2\,q(T_{\Qphase,t})(\rho_{\Qphase,t}+p_t).
 \label{eq:envelope}
\end{equation}
If the quark trace range shares even one trace value with the hadronic
one, including an occupied endpoint, then \(h=0\) on both.
\end{corollary}
\begin{proof}
Proposition~\ref{prop:surface} gives \(h=q=0\) on the hadronic branch, so
\(A_\Hphase=0\) and Eq.~\eqref{eq:interface} becomes
Eq.~\eqref{eq:envelope}. At any shared trace value, continuity gives
\(h=q=0\) on the quark branch. Uniqueness in Eq.~\eqref{eq:q-rho}
propagates the zero of \(q\) along that branch, including from an occupied
endpoint, and \(h\) keeps the value \(0\) there.
\end{proof}

The hadronic data of Section~\ref{sec:construction} have
\(C_\Hphase=(1/20)(11/10)^{5/2}\neq0\), so that branch cannot be continued
to zero density along the same polytrope. Within this model, a nonconstant
\(h\) compatible with a polytropic envelope must therefore be confined to
the inner phase. Keeping the quark slope \(q=1/20\) of
Eq.~\eqref{eq:bulkdata}, for example, Eq.~\eqref{eq:envelope} requires
\(\bar h(1.7)=21/100\) instead of \(1/10\), and a quintic extension from
\((\bar h,q,q_T)=(0,0,0)\) at \(\bar T=0.7\) again provides a \(C^2\)
function. A different envelope equation of state would require a new
reconstruction and new interface tests.

\subsection{Time-dependent matter}
\label{sec:dynamics}
\begin{proposition}[Dynamical restriction]
\label{prop:dynamics}
Assume (A1)--(A3) on an open region of spacetime, without (A4). Then
\begin{equation}
 q\,(1+3\cs)\,u^\mu\nabla_\mu\rho=0.
 \label{eq:timeprojection}
\end{equation}
Wherever \(q\neq0\), every fluid element keeps its density,
\(\dot\rho\equiv u^\mu\nabla_\mu\rho=0\), and the flow is
volume-preserving, \(\theta\equiv\nabla_\mu u^\mu=0\).
\end{proposition}
\begin{proof}
Since \(u^\nu(T_{\mu\nu}-\rho g_{\mu\nu})=0\), contracting
Eq.~\eqref{eq:covconstraint} with \(u^\nu\) leaves
\(q\,u^\nu\nabla_\nu(\rho+3p)=0\), which is Eq.~\eqref{eq:timeprojection}
for a barotrope. As \(1+3\cs\geq1\), \(q\neq0\) implies \(\dot\rho=0\).
The energy equation \(u^\nu\nabla^\mu T_{\mu\nu}=\dot\rho+w\theta=0\)
with \(w>0\) then gives \(\theta=0\).
\end{proof}

In a static background the projection~\eqref{eq:timeprojection} is
automatic, which is why it plays no role in Sections~\ref{sec:shared}
to~\ref{sec:construction}. Linearized about a static background with
\(q\neq0\), it reads \(\partial_t\Delta\rho=0\) for the Lagrangian density
perturbation \(\Delta\rho\). Every mode with nonzero frequency then has
\(\Delta\rho=0\) and, for a barotrope, \(\Delta p=\cs\Delta\rho=0\). The
compressional pressure response that drives the standard radial
pulsation problem \cite{Chandrasekhar1964} is absent. Static conservation
is therefore not a dynamical guarantee: the static constructions of
Sections~\ref{sec:shared} to~\ref{sec:construction} do not, by
themselves, define a framework for compressible perturbations with
separately conserved matter. For constant \(q\), that is, \(h\) linear in
\(T\), the conclusion that the energy density cannot vary was obtained in
Ref.~\cite{CarvalhoLobato2026}. The pulsation system of
Ref.~\cite{Pretel2021} is written in signature \((-+++)\) with
\(\mathcal L_m=-\rho\) and \(\Theta_{\mu\nu}=-2T_{\mu\nu}-\rho g_{\mu\nu}\),
which is the present closure. In arXiv:2105.07573v2, Eq.~(9) is the
covariant conservation constraint, Eq.~(44) is the perturbed radial
material-conservation equation, and Eqs.~(45) and~(52) give the density
and pressure perturbations. That system combines the perturbed
gravitational equations with radial material conservation and generally
allows \(\Delta\rho\neq0\). Our consistency test additionally imposes the time
projection~\eqref{eq:timeprojection}, which requires
\(q(1+3\cs)\,\partial_t\Delta\rho=0\) at linear order; for a mode
\(\propto e^{i\omega t}\) its residual is
\(i\omega\,u^t q(1+3\cs)\Delta\rho\). Compressive modes with
\(\omega\neq0\) therefore do not satisfy this additional condition where
\(h_T\neq0\). Even with bulk consistency secured, a hybrid-star pulsation
problem also needs interface conditions set by the phase-conversion
timescale, slow or rapid \cite{Pereira2018}; such dynamical conditions
would add to, not replace, the static traction
condition~\eqref{eq:interface}.

This restriction is tied to the closure. With
\(\Theta_{\mu\nu}=-2T_{\mu\nu}-pg_{\mu\nu}\) (\(\mathcal L_m=-p\)), the
same divergence gives
\begin{equation}
 w\,u_\nu u^\mu\nabla_\mu q+\frac q2\nabla_\nu(\rho-p)=0 .
 \label{eq:otherclosure}
\end{equation}
Its time projection is the equation from which conservative cosmologies
were reconstructed \cite{Alvarenga2013,Chakraborty2013}: for
\(p=\omega\rho\) with \(\omega\neq-1,\pm1/3\) it gives
\(h=\alpha T^{(1+3\omega)/(2+2\omega)}+\beta\); the primitive is
logarithmic at \(\omega=-1/3\), and the trace is constant at
\(\omega=1/3\). The power is written for \(T>0\); on a negative-trace
interval a real branch is expressed using \(|T|\), with signs absorbed
in the integration constant. Its spatial projection forces \(q=0\) wherever
\(\rho-p\) varies in space, as found in
Refs.~\cite{Bertini2023,CarvalhoLobato2026}. For branches along which
\(\rho-p\) varies, the two closures therefore constrain different sectors
where \(q\neq0\): the energy-density closure forbids comoving density
evolution, while the pressure closure forbids static stratification. A
stiff branch with \(\rho-p\) constant escapes the spatial obstruction.
With the prescription of Ref.~\cite{Ky2026}, by contrast, the time
projection vanishes identically for barotropic flows that conserve
baryon number.

\subsection{Closure and relation to previous work}
\label{sec:previous}
The identification \(\Theta_{\mu\nu}=-2T_{\mu\nu}+\rho g_{\mu\nu}\) is an
assumption, not a consequence of writing \(\mathcal L_m=\rho\).
Refs.~\cite{Haghani2024,Ky2026,CarvalhoLobato2026} reexamine the metric
variation of perfect-fluid stress and its consequences for conservation;
Ref.~\cite{Ky2026} proposes a prescription independent of
\(\mathcal L_m\) and explicitly revisits conservative stellar models.
Derivations from the full fluid actions of Brown and Schutz
\cite{Brown1993,Schutz1970} give trace-coupling terms that differ from
those of the on-shell prescriptions \cite{Boehmer2025}. In Brown's
formulation, for \(f=R+h(T)\), the extra term is proportional to
\(n\,\partial T/\partial n\) times the spatial projector, the structure of
the prescription of Ref.~\cite{Ky2026}, and the \(h\)-dependent terms can
be regrouped into a modified matter source \cite{Boehmer2025}. Whether
such a regrouping removes the physical content of \(h\) has been debated
\cite{Fisher2019,HarkoMoraes2020,Fisher2020}. Our results concern the
specified field equation~\eqref{eq:field} and a separately conserved
material tensor with a prescribed equation of state; they make no claim
of action-level equivalence or uniqueness.

For algebraic effective sources of this kind, a shell-free interface
still requires continuity of the effective normal stress. The bulk
constraint, however, depends on the closure: it may give a regular
first-order linear equation for \(q\), an algebraic restriction, or a
degenerate identity. For example, the static spatial projection of
Eq.~\eqref{eq:otherclosure} is \(q\,\partial_i(\rho-p)=0\), with no
derivative of \(q\). A shared-trace criterion of the form of
Lemma~\ref{lem:shared} applies only when the closure supplies a regular,
nondegenerate equation \(q_T=-Fq\) on each branch. Its coefficient and
Propositions~\ref{prop:obstruction}, \ref{prop:surface},
and~\ref{prop:dynamics} must be rederived for each closure; for the
prescription of Ref.~\cite{Ky2026} the time projection vanishes
identically (Section~\ref{sec:dynamics}). Every proposition here is
conditional on Eq.~\eqref{eq:field} and on separate conservation.

The bulk conservation condition is that of
Refs.~\cite{Santos2019,Carvalho2020,Pretel2021}, which use the same
closure and impose the radial component. That the reconstructed function
depends on the equation of state was noted in
Refs.~\cite{Santos2019,Lobato2026}. Lemma~\ref{lem:shared}, itself a
consequence of uniqueness for linear ordinary differential equations, and
Proposition~\ref{prop:obstruction} turn this dependence into an
obstruction for two phases that share trace values. That the trace may
jump when \(f_R\) is constant follows from the general junction analyses
\cite{Rosa2021,RosaRubiera2022,Rosa2024}, and
Proposition~\ref{prop:interface} is the Darmois condition applied to the
effective stress of that subclass. What we add is its identification as
the condition that material equilibrium leaves open at a first-order
transition, together with the construction of
Proposition~\ref{prop:construction}. The \(\rho^{-1/2}\) behaviour behind
Proposition~\ref{prop:surface} is already visible in the polytropic
reconstructions of Refs.~\cite{Santos2019,Pretel2021} as
\(\rho\to0\); its consequence for the regularity of \(h\) at \(T=0\) is
not discussed there. Proposition~\ref{prop:dynamics} extends the
constant-\(q\) result of Ref.~\cite{CarvalhoLobato2026} to arbitrary
\(h\) under the present closure; projections of the conservation
constraint along and orthogonal to \(u^\mu\) were used for the closure
\(\mathcal L_m=-p\) in Ref.~\cite{Bertini2023}.

\subsection{Status of the results}
\label{sec:status}
\emph{Demonstrated}, under the hypotheses stated with each result:
Lemmas~\ref{lem:shared} and~\ref{lem:turning},
Propositions~\ref{prop:obstruction}--\ref{prop:dynamics}, and
Corollary~\ref{cor:envelope}. \emph{Illustrated} with synthetic data:
the overlap of Fig.~\ref{fig:shared}, the interface of
Fig.~\ref{fig:interface}, and the construction of
Fig.~\ref{fig:construction}. \emph{Open}: a complete hybrid configuration
with one fixed \(h\), including a regular center, the envelope, and the
exterior; a treatment of perturbations compatible with
Proposition~\ref{prop:dynamics}; an action-level foundation for the
closure; and realistic equations of state. For the single function
\(h\), these results locate where a nonconstant part can reside: in an
inner phase, on trace values separated from those of a polytropic
envelope, and in static matter.

\section{Conclusions}
\label{sec:conclusions}
We asked whether one function \(h(T)\) can serve two matter phases
separated by a sharp transition in conservative \(f(R,T)=R+h(T)\) gravity,
for static backgrounds, the \(\mathcal L_m=\rho\) closure, and a
shell-free Maxwell interface. Beyond the single-phase reconstructions,
two conditions are necessary: the phases must require the same evolution
of \(q=h'(T)\) at every shared trace value, and \(q(\rho+p)-h/2\) must be
continuous at the interface. Material equilibrium does not supply the
second condition.

For a hadronic polytrope and an affine quark phase, an open overlap of
the occupied trace ranges leaves only a constant \(h\). Separated ranges
admit a nonconstant \(C^2\) function that satisfies both conditions at a
single interface, as the explicit construction shows.

This possibility is local. A polytropic envelope that reaches zero density
with \(h\) regular at \(T=0\) carries no nonconstant part of \(h\), and
wherever \(h'(T)\neq0\), separate conservation fixes the density of every
fluid element. A hybrid configuration with nonconstant \(h\) and such an
envelope must therefore confine that part to the inner phase. Whether a
complete configuration of this kind exists is a global boundary-value
problem with one fixed \(h\).

\appendix
\section{Component sign audit}
\label{app:sign}
For the conventions used here, \(T^\mu{}_{\nu}=\operatorname{diag}
(\rho,-p,-p,-p)\). Raising an index in Eq.~\eqref{eq:field} gives
\begin{equation}
 G^t{}_t=8\pi\rho+h/2,\qquad
 G^r{}_r=-8\pi p+h/2-q(\rho+p).
\end{equation}
Consequently the radial pressure source is
\(8\pi p_{\rm eff}=8\pi p-h/2+q(\rho+p)\). This passes a direct Bianchi
check: \(\rho_{\rm eff}+p_{\rm eff}=w(1+q/8\pi)\) and
\(p_{\rm eff}'=(1+q/8\pi)p'\) on a conservative branch, so effective
equilibrium follows from \(p'=-w\phi'/2\).

The accessible preprint arXiv:1803.07719v3 displays a component term
\(-h_T(p-\rho)\) in its Eq.~(11b). With the preprint's own covariant
Eq.~(3), \(\mathcal L_m=\rho\), and signature \((+---)\), the radial
source instead contains \(+h_T(\rho+p)\). We therefore use the covariant
equation as the defining prescription. This observation is restricted to
the checked preprint; it does not assert that the published numerical
implementation used the same transcription, nor does it reassess its
mass--radius results.

\section{Single-phase formulas and units}
\label{app:formulas}
For the affine phase, with \(b=(1-s)/(2+2s)\),
\begin{equation}
 h_\Qphase=B_\Qphase+\frac{C_\Qphase\,(1-3s)}{(1+s)\,b}\,w_\Qphase^{b},
 \qquad s\neq1,\;1/3.
 \label{eq:hQ}
\end{equation}
At \(s=1\) the primitive is \(h_\Qphase=B_\Qphase-C_\Qphase
\ln(w_\Qphase/w_a)\), where \(w_a>0\) makes the argument dimensionless. At
\(s=1/3\), Eq.~\eqref{eq:constanttrace} replaces both formulas. For the
polytrope, the primitive is kept as the definite quadrature
\eqref{eq:hsol}, including \(B_\Hphase\).

Integrating the first law~\eqref{eq:firstlaw} gives
\begin{align}
 n_\Hphase&=N_\Hphase\,\rho\left(1+K\rho^{\Gamma-1}\right)^{-1/(\Gamma-1)},
 \label{eq:nH}\\
 n_\Qphase&=N_\Qphase\,w_\Qphase^{1/(1+s)},
 \label{eq:nQ}
\end{align}
so that \(n_\Qphase/n_\Hphase=(\rho_\Qphase+p_t)/(\rho_\Hphase+p_t)\) at
coexistence. The constants \(N_i>0\) carry units and encode the
thermodynamic normalization.

For the soft pair with the data~\eqref{eq:bulkdata},
\(\Gamma=2\), \(\bar K=1/10\), \(s=1/5\), and \(\bar\rho_0=3/2\), the bulk
functions are
\begin{align}
 q_\Hphase&=\frac{1}{20}\left(\frac{11}{10}\right)^{5/2}
    \bar\rho^{-1/2}\left(1+\frac{\bar\rho}{10}\right)^{-5/2},\qquad
 \bar h_\Hphase=\int_1^{\bar\rho}q_\Hphase(z)\left(1-\frac{3z}{5}\right)\dd z,
 \label{eq:exH}\\
 q_\Qphase&=\frac{1}{20}\left(\frac{10\bar w_\Qphase}{21}\right)^{-2/3},
 \qquad
 \bar h_\Qphase=\frac{1}{10}+\frac{1}{20}\left(\frac{21}{10}\right)^{2/3}
   \left[\bar w_\Qphase^{1/3}-\left(\frac{21}{10}\right)^{1/3}\right],
 \label{eq:exQ}
\end{align}
with \(\bar w_\Qphase=6\bar\rho/5-3/10\). The baryon normalizations are
\(\bar n_\Hphase=(11/10)\,\bar\rho/(1+\bar\rho/10)\) and
\(\bar n_\Qphase=(21/11)\,(10\bar w_\Qphase/21)^{5/6}\).

In geometrized units, \(\rho\), \(p\), \(T\), and \(h\) have dimension
\(L^{-2}\); \(q\) is dimensionless; \(h_{TT}\) and \(F\) have dimension
\(L^2\); \(K\) has dimension \(L^{2(\Gamma-1)}\); and \(s\) is
dimensionless. The amplitudes \(C_i\) carry the units that compensate the
density powers in Eqs.~\eqref{eq:qH} and~\eqref{eq:qQ}.

\acknowledgments
During the preparation of this manuscript, the authors used Claude
(Anthropic) to support restructuring and editing of the text, checking
derivations, and the preparation of the figures. But, we would like to
stress this, the main results and their derivations originate in the
authors' own earlier work. The authors critically revised all content,
checked every cited source against the original and take full
responsibility for the article.


\begin{thebibliography}{99}
\bibitem{Harko2011}
T. Harko, F. S. N. Lobo, S. Nojiri and S. D. Odintsov,
\emph{\(f(R,T)\) gravity}, Phys. Rev. D \textbf{84}, 024020 (2011).
\href{https://doi.org/10.1103/PhysRevD.84.024020}{doi:10.1103/PhysRevD.84.024020};
\href{https://arxiv.org/abs/1104.2669}{arXiv:1104.2669}.

\bibitem{Alvarenga2013}
F. G. Alvarenga, A. de la Cruz-Dombriz, M. J. S. Houndjo, M. E. Rodrigues
and D. S\'aez-G\'omez,
\emph{Dynamics of scalar perturbations in \(f(R,T)\) gravity},
Phys. Rev. D \textbf{87}, 103526 (2013).
\href{https://doi.org/10.1103/PhysRevD.87.103526}{doi:10.1103/PhysRevD.87.103526};
\href{https://arxiv.org/abs/1302.1866}{arXiv:1302.1866}.

\bibitem{Chakraborty2013}
S. Chakraborty,
\emph{An alternative \(f(R,T)\) gravity theory and the dark energy problem},
Gen. Relativ. Gravit. \textbf{45}, 2039 (2013).
\href{https://doi.org/10.1007/s10714-013-1577-y}{doi:10.1007/s10714-013-1577-y};
\href{https://arxiv.org/abs/1212.3050}{arXiv:1212.3050}.

\bibitem{Santos2019}
S. I. dos Santos Jr., G. A. Carvalho, P. H. R. S. Moraes,
C. H. Lenzi and M. Malheiro,
\emph{A conservative energy-momentum tensor in the \(f(R,T)\) gravity
and its implications for the phenomenology of neutron stars},
Eur. Phys. J. Plus \textbf{134}, 398 (2019).
\href{https://doi.org/10.1140/epjp/i2019-12830-8}{doi:10.1140/epjp/i2019-12830-8};
\href{https://arxiv.org/abs/1803.07719}{arXiv:1803.07719}.

\bibitem{Pretel2021}
J. M. Z. Pretel, S. E. Jor\'as, R. R. R. Reis and J. D. V. Arba\~nil,
\emph{Neutron stars in \(f(R,T)\) gravity with conserved energy-momentum
tensor: Hydrostatic equilibrium and asteroseismology},
JCAP \textbf{08}, 055 (2021).
\href{https://doi.org/10.1088/1475-7516/2021/08/055}{doi:10.1088/1475-7516/2021/08/055};
\href{https://arxiv.org/abs/2105.07573}{arXiv:2105.07573}.

\bibitem{Carvalho2020}
G. A. Carvalho, S. I. dos Santos Jr., P. H. R. S. Moraes and M. Malheiro,
\emph{Strange stars in energy-momentum-conserved \(f(R,T)\) gravity},
Int. J. Mod. Phys. D \textbf{29}, 2050075 (2020).
\href{https://doi.org/10.1142/S0218271820500753}{doi:10.1142/S0218271820500753};
\href{https://arxiv.org/abs/1911.02484}{arXiv:1911.02484}.

\bibitem{Lobato2026}
R. V. Lobato, G. A. Carvalho, C. E. C. Monta\~na and J. F. Rodriguez-Ruiz,
\emph{Neutron stars in a conservative \(f(R,T)\) gravity},
preprint (2026).
\href{https://arxiv.org/abs/2605.09206}{arXiv:2605.09206}.

\bibitem{Rosa2021}
J. L. Rosa, \emph{Junction conditions and thin shells in perfect-fluid
\(f(R,T)\) gravity}, Phys. Rev. D \textbf{103}, 104069 (2021).
\href{https://doi.org/10.1103/PhysRevD.103.104069}{doi:10.1103/PhysRevD.103.104069};
\href{https://arxiv.org/abs/2103.11698}{arXiv:2103.11698}.

\bibitem{RosaRubiera2022}
J. L. Rosa and D. Rubiera-Garcia,
\emph{Junction conditions of Palatini \(f(\mathcal R,T)\) gravity},
Phys. Rev. D \textbf{106}, 064007 (2022).
\href{https://doi.org/10.1103/PhysRevD.106.064007}{doi:10.1103/PhysRevD.106.064007};
\href{https://arxiv.org/abs/2204.12944}{arXiv:2204.12944}.

\bibitem{Rosa2024}
J. L. Rosa,
\emph{Junction conditions in gravity theories with extra scalar degrees of
freedom}, Phys. Rev. D \textbf{109}, 064018 (2024).
\href{https://doi.org/10.1103/PhysRevD.109.064018}{doi:10.1103/PhysRevD.109.064018};
\href{https://arxiv.org/abs/2311.11752}{arXiv:2311.11752}.

\bibitem{Bertini2023}
N. R. Bertini and H. Velten,
\emph{Fully conservative \(f(R,T)\) gravity and Solar System constraints},
Phys. Rev. D \textbf{107}, 124005 (2023).
\href{https://doi.org/10.1103/PhysRevD.107.124005}{doi:10.1103/PhysRevD.107.124005};
\href{https://arxiv.org/abs/2303.09699}{arXiv:2303.09699}.

\bibitem{Haghani2024}
Z. Haghani, T. Harko and S. Shahidi,
\emph{The first variation of the matter energy-momentum tensor with
respect to the metric, and its implications on modified gravity theories},
Phys. Dark Univ. \textbf{44}, 101448 (2024).
\href{https://doi.org/10.1016/j.dark.2024.101448}{doi:10.1016/j.dark.2024.101448};
\href{https://arxiv.org/abs/2301.12133}{arXiv:2301.12133}.

\bibitem{Ky2026}
P. V. Ky, \emph{Metric variation of the energy-momentum tensor of a perfect
fluid and its applications to cosmology and neutron stars},
Eur. Phys. J. C \textbf{86}, 912 (2026).
\href{https://doi.org/10.1140/epjc/s10052-026-16185-y}{doi:10.1140/epjc/s10052-026-16185-y};
\href{https://arxiv.org/abs/2610.05763}{arXiv:2610.05763}.

\bibitem{CarvalhoLobato2026}
G. A. Carvalho and R. V. Lobato,
\emph{Prescription dependence of the perfect-fluid Lagrangian and
conservation constraints in non-minimally coupled gravity},
Eur. Phys. J. Plus \textbf{141}, 1083 (2026).
\href{https://doi.org/10.1140/epjp/s13360-026-08342-9}{doi:10.1140/epjp/s13360-026-08342-9}.

\bibitem{Brown1993}
J. D. Brown, \emph{Action functionals for relativistic perfect fluids},
Class. Quantum Grav. \textbf{10}, 1579 (1993).
\href{https://doi.org/10.1088/0264-9381/10/8/017}{doi:10.1088/0264-9381/10/8/017}; \href{https://arxiv.org/abs/gr-qc/9304026}{arXiv:gr-qc/9304026}.

\bibitem{Schutz1970}
B. F. Schutz, \emph{Perfect fluids in general relativity: velocity
potentials and a variational principle},
Phys. Rev. D \textbf{2}, 2762 (1970).
\href{https://doi.org/10.1103/PhysRevD.2.2762}{doi:10.1103/PhysRevD.2.2762}.

\bibitem{Boehmer2025}
C. G. B\"ohmer and E. Al-Nasrallah,
\emph{Consistent energy-momentum trace couplings of fluids},
Phys. Rev. D \textbf{112}, 124073 (2025).
\href{https://doi.org/10.1103/6wrh-lg9s}{doi:10.1103/6wrh-lg9s}; \href{https://arxiv.org/abs/2509.24843}{arXiv:2509.24843}.

\bibitem{BarrosSa2026}
N. Barros e S\'a, M. A. S. Pinto and T. Trindade,
\emph{Covariant conservation laws, local invariance and Noether's second
theorem}, Eur. Phys. J. C \textbf{86}, 612 (2026).
\href{https://doi.org/10.1140/epjc/s10052-026-15850-6}{doi:10.1140/epjc/s10052-026-15850-6}; \href{https://arxiv.org/abs/2506.14454}{arXiv:2506.14454}.

\bibitem{Glendenning1992}
N. K. Glendenning,
\emph{First-order phase transitions with more than one conserved charge:
consequences for neutron stars}, Phys. Rev. D \textbf{46}, 1274 (1992).
\href{https://doi.org/10.1103/PhysRevD.46.1274}{doi:10.1103/PhysRevD.46.1274}.

\bibitem{Bedaque2015}
P. Bedaque and A. W. Steiner, \emph{Sound velocity bound and neutron
stars}, Phys. Rev. Lett. \textbf{114}, 031103 (2015).
\href{https://doi.org/10.1103/PhysRevLett.114.031103}{doi:10.1103/PhysRevLett.114.031103}; \href{https://arxiv.org/abs/1408.5116}{arXiv:1408.5116}.

\bibitem{Garibay2026}
I. Garibay, C. Ecker and L. Rezzolla,
\emph{General gravitational properties of neutron stars: curvature
invariants, binding energy, and trace anomaly},
Phys. Rev. D \textbf{113}, 083028 (2026).
\href{https://doi.org/10.1103/tkm4-3pcr}{doi:10.1103/tkm4-3pcr}; \href{https://arxiv.org/abs/2601.07931}{arXiv:2601.07931}.

\bibitem{Jimenez2024}
J. C. Jim\'enez, L. Lazzari and V. P. Gon\c{c}alves,
\emph{How the QCD trace anomaly behaves at the core of twin stars?},
Phys. Rev. D \textbf{110}, 114014 (2024).
\href{https://doi.org/10.1103/PhysRevD.110.114014}{doi:10.1103/PhysRevD.110.114014}; \href{https://arxiv.org/abs/2408.11614}{arXiv:2408.11614}.

\bibitem{Alford2013}
M. G. Alford, S. Han and M. Prakash,
\emph{Generic conditions for stable hybrid stars},
Phys. Rev. D \textbf{88}, 083013 (2013).
\href{https://doi.org/10.1103/PhysRevD.88.083013}{doi:10.1103/PhysRevD.88.083013};
\href{https://arxiv.org/abs/1302.4732}{arXiv:1302.4732}.

\bibitem{Darmois1927}
G. Darmois, \emph{Les \'equations de la gravitation einsteinienne},
M\'emorial des Sciences Math\'ematiques, Fascicule 25
(Gauthier-Villars, Paris, 1927).

\bibitem{Israel1966}
W. Israel,
\emph{Singular hypersurfaces and thin shells in general relativity},
Nuovo Cimento B \textbf{44}, 1 (1966).
\href{https://doi.org/10.1007/BF02710419}{doi:10.1007/BF02710419}.

\bibitem{Chandrasekhar1964}
S. Chandrasekhar,
\emph{The dynamical instability of gaseous masses approaching the
Schwarzschild limit in general relativity},
Astrophys. J. \textbf{140}, 417 (1964).
\href{https://doi.org/10.1086/147938}{doi:10.1086/147938}.

\bibitem{Pereira2018}
J. P. Pereira, C. V. Flores and G. Lugones,
\emph{Phase transition effects on the dynamical stability of hybrid
neutron stars}, Astrophys. J. \textbf{860}, 12 (2018).
\href{https://doi.org/10.3847/1538-4357/aabfbf}{doi:10.3847/1538-4357/aabfbf}; \href{https://arxiv.org/abs/1706.09371}{arXiv:1706.09371}.

\bibitem{Fisher2019}
S. B. Fisher and E. D. Carlson, \emph{Reexamining \(f(R,T)\) gravity},
Phys. Rev. D \textbf{100}, 064059 (2019).
\href{https://doi.org/10.1103/PhysRevD.100.064059}{doi:10.1103/PhysRevD.100.064059}.

\bibitem{HarkoMoraes2020}
T. Harko and P. H. R. S. Moraes,
\emph{Comment on ``Reexamining \(f(R,T)\) gravity''},
Phys. Rev. D \textbf{101}, 108501 (2020).
\href{https://doi.org/10.1103/PhysRevD.101.108501}{doi:10.1103/PhysRevD.101.108501}; \href{https://arxiv.org/abs/2003.08107}{arXiv:2003.08107}.

\bibitem{Fisher2020}
S. B. Fisher and E. D. Carlson,
\emph{Reply to ``Comment on `Reexamining \(f(R,T)\) gravity'\,''},
Phys. Rev. D \textbf{101}, 108502 (2020).
\href{https://doi.org/10.1103/PhysRevD.101.108502}{doi:10.1103/PhysRevD.101.108502}; \href{https://arxiv.org/abs/2004.02934}{arXiv:2004.02934}.
\end{thebibliography}
\end{document}